%% file: fr3-isac-mag.tex
\documentclass[lettersize,journal]{IEEEtran}
\usepackage{amsmath,amsfonts}
\usepackage{algorithmic}
\usepackage{algorithm}
\usepackage{array}
\usepackage[caption=false,font=normalsize,labelfont=sf,textfont=sf]{subfig}
\usepackage{textcomp}
\usepackage{stfloats}
\usepackage{url}
\usepackage{verbatim}
\usepackage{graphicx}
\usepackage{soul}
\usepackage{textcomp}
\usepackage{stfloats}
\usepackage{float}
\usepackage{url}
\usepackage{pgf}
\usepackage{verbatim}
\usepackage{graphicx}
\usepackage{cite}
\usepackage{makecell}
\usepackage{amsmath}
\usepackage{tikz}
\usepackage{xcolor}
\usetikzlibrary{arrows.meta}
\usepackage{pgfplots}
\usepackage{standalone}
\usepackage{xcolor}
\usepackage{cite}
\usepackage{soul}
\usepackage{tabularx,booktabs,enumitem}
\usepackage{xcolor}
\sethlcolor{yellow}  
\usepackage[nolist]{acronym}
\usepackage{siunitx}
\DeclareMathOperator{\dB}{dB}
\DeclareMathOperator{\Gbps}{Gbps}
\DeclareMathOperator{\Mbps}{Mbps}
\DeclareMathOperator{\bpd}{bits/DoF}
\DeclareMathOperator{\MHz}{MHz}
\DeclareMathOperator{\GHz}{GHz}

\DeclareMathOperator{\meters}{m}
\DeclareMathOperator{\kilometers}{km}
\DeclareMathOperator{\nanoseconds}{ns}

\usepackage{titlesec}
\usepackage{multirow}

 \usepackage{xcolor,cite,etoolbox}
\makeatletter 
\pretocmd\@bibitem{\color{black}\csname keycolor#1\endcsname}{}{\fail}
\newcommand\citecolor[1]{\@namedef{keycolor#1}{\color{black}}}
\makeatother

\input{acronyms.tex}

\usepackage{fancyhdr}
\fancypagestyle{firststyle}{
	\fancyhf{}
	\fancyhead[L]{A. Bazzi, R. Bomfin, M. Mezzavilla, S. Rangan, T. S. Rappaport and M. Chafii, ``Upper Mid-Band Spectrum for 6G: \\Vision, Opportunity and Challenges'', accepted in \textit{IEEE Communications Magazine}, 2025.}

}

\def\BibTeX{{\rm B\kern-.05em{\sc i\kern-.025em b}\kern-.08em
    T\kern-.1667em\lower.7ex\hbox{E}\kern-.125emX}}

\begin{document}

\title{Upper Mid-Band Spectrum for 6G: \\Vision, Opportunity and Challenges}

\author{Ahmad Bazzi,
Roberto Bomfin,
Marco Mezzavilla,
Sundeep Rangan,
Theodore S. Rappaport,
Marwa Chafii
\thanks{Ahmad Bazzi, Roberto Bomfin, and Marwa Chafii are with the Engineering Division, New York University Abu Dhabi (NYUAD), 129188, UAE
(email: \{ahmad.bazzi,roberto.bomfin, marwa.chafii\}@nyu.edu)}
\thanks{Ahmad Bazzi, Marwa Chafii, Sundeep Rangan, and Theodore Rappaport are with NYU WIRELESS, NYU Tandon School of Engineering, Brooklyn, 11201, NY, USA (email: \{srangan,tsr\}@nyu.edu).}
\thanks{Marco Mezzavilla is with the Telecommunications Engineering at the Department of Electronics, Information, and Bioengineering (DEIB) at Politecnico di Milano, Milan, Italy.
(email: marco.mezzavilla@polimi.it).}
}

\markboth{}%
{Shell \MakeLowercase{\textit{et al.}}: A Sample Article Using IEEEtran.cls for IEEE Journals}


\maketitle
\thispagestyle{firststyle}

\begin{abstract}
Driven by the pursuit of gigabit-per-second data speeds for future 6G mobile networks, in addition to the support of sensing and artificial intelligence applications, the industry is expanding beyond crowded sub-6 GHz bands with innovative new spectrum allocations. 
In this paper, we chart a compelling vision for 6G within the frequency range 3 (FR3) spectrum, i.e. $7.125$-$24.25$ $\GHz$, by delving into its key enablers and addressing the multifaceted challenges that lie ahead for these new frequency bands.
Here we highlight the physical properties of this never-before used spectrum \textcolor{black}{for cellular} by reviewing recent channel measurements for outdoor and indoor environments, including path loss, delay and angular spreads, and material penetration loss, all which offer insights that underpin future 5G/6G wireless communication designs. Building on the fundamental knowledge of the channel properties, we explore FR3 spectrum agility strategies that balance coverage and capacity tradeoffs, while examining coexistence with incumbent systems, such as satellites, radio astronomy, and earth exploration. 
Moreover, we discuss the potential of massive multiple-input multiple-output \textcolor{black}{technologies}, \textcolor{black}{challenges for commercial deployment, and potential solutions for  FR3, including} multiband sensing for FR3 integrated sensing and communications.
Finally, we outline 6G standardization features that are likely to emerge from 3GPP radio frame innovations and open radio access network developments.
\end{abstract}


\section{Introduction}
\label{sec:intro}
\input{sections/introduction.tex}

\section{FR3 Channel Characteristics} 
\label{sec:channel-char}

\input{sections/channel-characteristics.tex}
\begin{figure}[t!]
    \centering
    \includegraphics[scale=0.15]{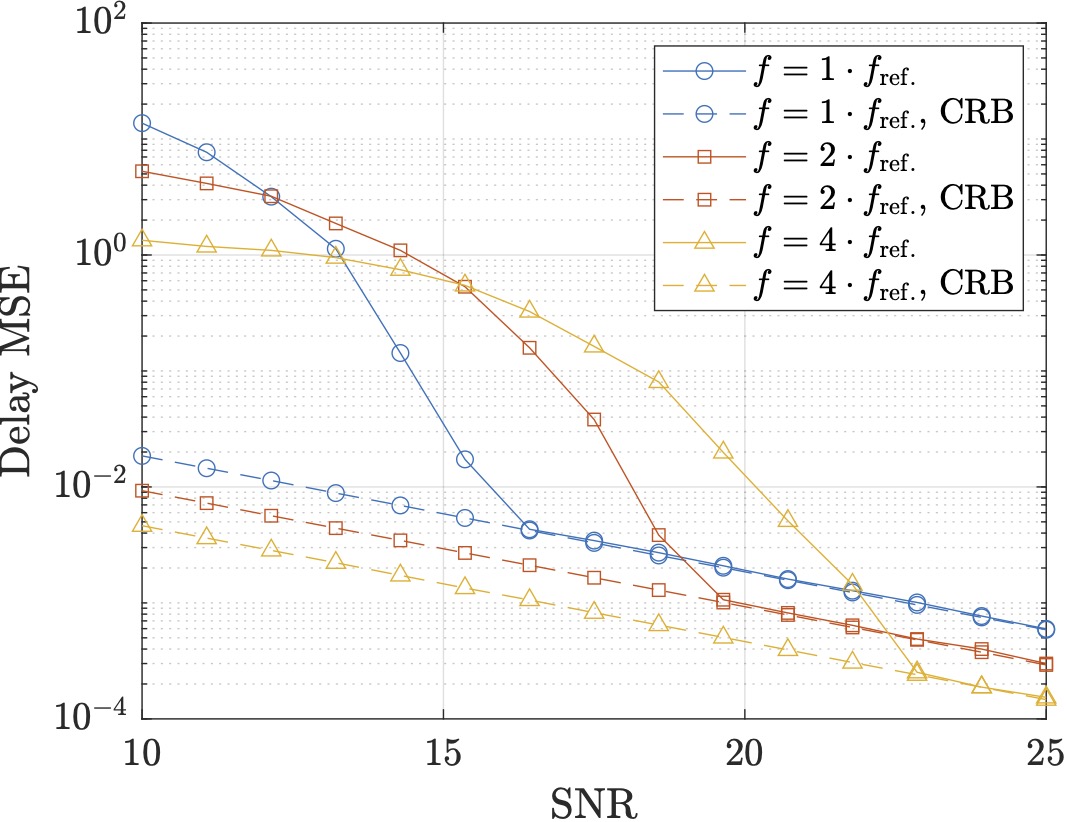}
    \caption{Multi-frequency ISAC trade-offs between low and high frequencies.}
    \label{fig:multi_band_sensing}
\end{figure}

\begin{figure}[t!]
    \centering
    \includegraphics[scale=0.15]{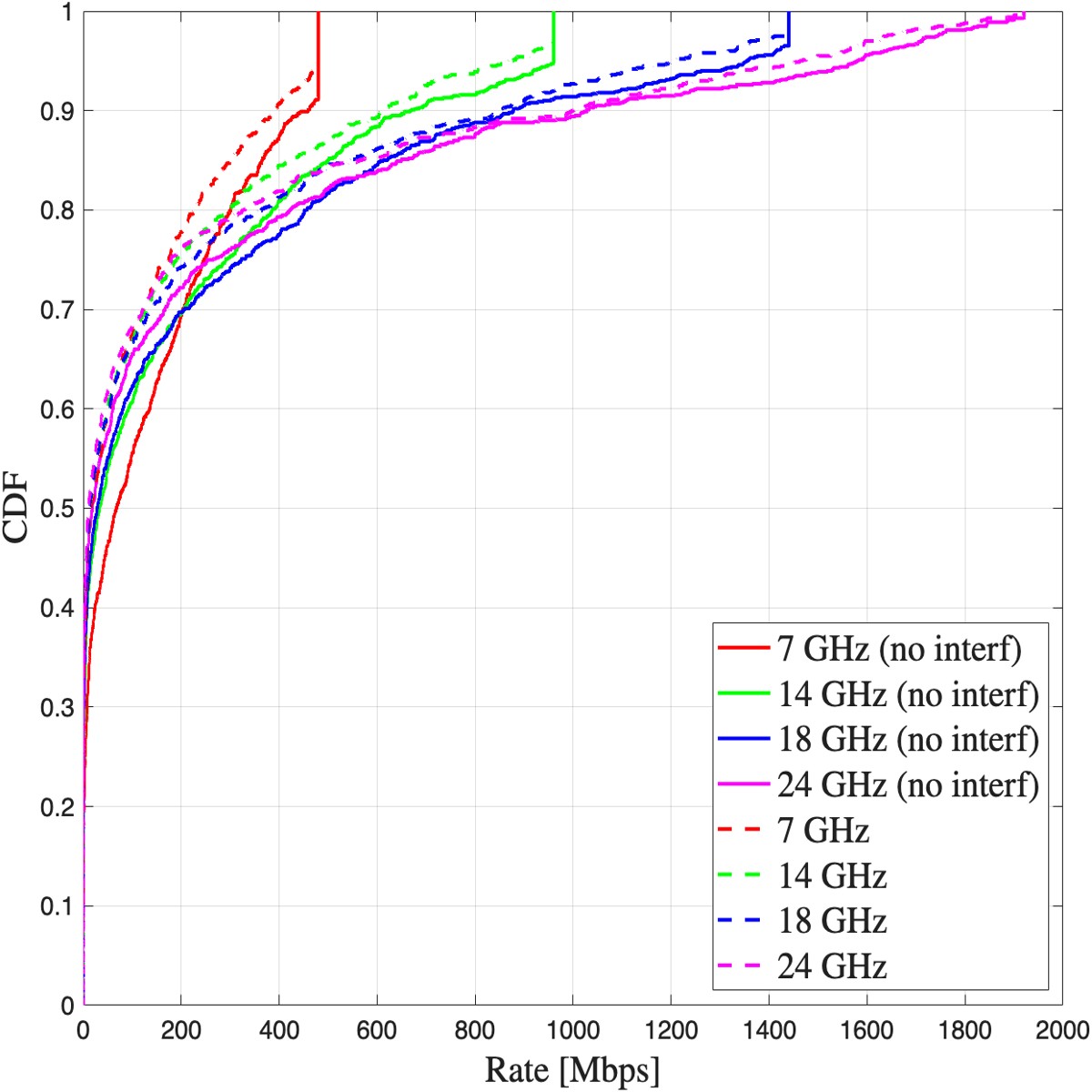}
    \caption{\textcolor{black}{Rate-coverage analysis over $7$ GHz, $14$ GHz, $18$ GHz and $24$ GHz.}}
    \label{fig:rate_coverage}
\end{figure}

\textcolor{black}{\input{actions/consVspros.tex}}

\section{\textcolor{black}{ITU enhancements via FR3}}
\label{sec:itu-enhancements}
\textcolor{black}{\input{actions/itu-enhancements-section}}


\input{actions/compare-table.tex}

\section{\textcolor{black}{Challenges for commercial deployment}}
\label{sec:challenges-commercial}
\textcolor{black}{\input{sections/commercial-challenges}}
%

\section{\textcolor{black}{Open Questions \& Potential Solutions}}
\label{sec:open-questions}
\textcolor{black}{\input{actions/open-questions}}

\section{Conclusions}
\label{sec:conclusions}
\input{sections/conclusions}

\bibliographystyle{IEEEtran}
\bibliography{refs}

\vfill

\end{document}

%% file: acronyms.tex
\begin{acronym}
	\acro{LoS}{line of sight}
        \acro{CSI}{channel state information}
	\acro{NLoS}{non line of sight}
	\acro{InH}{indoor hotspot}
	\acro{InF}{indoor factory}
	\acro{UMi}{urban microcell}
	\acro{SHF}{super high frequency}
	\acro{DS}{delay spread}
	\acro{RMS}{root mean square}
	\acro{AS}{angular spread}
        \acro{DSS}{dynamic spectrum sharing}
	\acro{3GPP}{$3^{\text{rd}}$ Generation Partnership Program}
    \acro{NGJ-MB}{next generation jammer mid-band}
	\acro{RF}{radio frequency}
	\acro{ACLR}{adjacent channel leakage power ratio}
        \acro{FCC}{Federal Communications Commission}
        \acro{NTIA}{National Telecommunications and Information Administration}
         \acro{CFO}{carrier frequency offset}
         \acro{CRB}{Cram{\'{e}}r-Rao bound}
         \acro{AWGN}{additive white Gaussian noise}
         \acro{FFT}{fast Fourier transform}
         \acro{CP}{cyclic prefix}
         \acro{MSE}{mean squared error}
         \acro{MRC}{maximum ratio combining}
         \acro{CPE}{customer-premises equipment}
         \acro{TRP}{transmission and reception point}
         \acro{PSD}{power spectral density}
         \acro{SE}{spectral efficiency}
         \acro{PN}{phase noise}
         \acro{AR}{augmented reality}
         \acro{XR}{extended reality}
         \acro{ISAC}{integrated sensing and communications}
         \acro{BS}{base station}
         \acro{FR1}{frequency range~1}
         \acro{FR2}{frequency range~2}
         \acro{FR3}{frequency range~3}
         \acro{PLE}{pathloss exponent}
         \acro{mmWave}{millimeter wave}
         \acro{gNB}{gNodeB}
         \acro{NR}{new radio}
         \acro{UE}{user equipment}
         \acro{MC}{mutual coupling}
         \acro{SINR}{signal-to-interference-plus-noise ratio}
         \acro{ITU}{International Telecommunication Union}
         \acro{MIMO}{multiple-input and multiple-output}
         \acro{ADC}{analog-to-digital converter}
         \acro{DAC}{digital-to-analog converter}
         \acro{PA}{power amplifier}
         \acro{O-RU}{open radio unit}
         \acro{DoF}{degrees-of-freedom}
         \acro{UL}{uplink}
         \acro{DL}{downlink}
         \acro{SHO}{soft handover}
         \acro{CDMA}{code-division multiple access}
         \acro{VR}{virtual reality}
         \acro{QAM}{quadrature amplitude modulation}
         \acro{LTE}{long-Term Evolution}
         \acro{SNR}{signal-to-noise ratio}
         \acro{RMT}{random matrix theory}
         \acro{ELAA}{extremely large aperture array}
         \acro{NF}{near-field}
         \acro{FF}{far-field}
         \acro{RT}{ray tracing}
         \acro{AI}{artificial intelligence}
         \acro{RT}{real-time}
         \acro{RAN}{radio access network}
         \acro{O-RAN}{open radio access network}
         \acro{RIC}{RAN intelligent controller}
         \acro{EC}{ergodic capacity}
         \acro{AoA}{angle-of-arrival}
         \acro{SAR}{synthetic aperture radar}
         \acro{RIS}{reconfigurable intelligent surface} 
         \acro{ILFD}{injection-locked frequency divider}
         \acro{TSPC}{true single-phase clock}
         \acro{ET}{extended target}
         \acro{UWB}{ultra-wide band}
         \acro{SnC}[S\&C]{sensing and communication}
         \acro{SB}{spectrum block}
         \acro{mMIMO}{massive MIMO}
         \acro{NB}{narrowband}
         \acro{EIRP}{equivalent isotropic radiated power}
         \acro{TDoc}{technical document}
         \acro{RAN}{radio access network}
         \acro{WG1}{working group}
         \acro{CIF}{close-in free space model with a frequency-weighted path loss exponent}
         \acro{TSG}{technical specification group}
         \acro{ASA}{angular spread of arrival}
         \acro{IEEE}{Institute of Electrical and Electronics Engineers}
         \acro{IMT}{International Mobile Telecommunication}
         \acro{CRLB}{Cram\'er-Rao lower bound}
         \acro{RSS}{received signal strength}
         \acro{DMC}{dense multipath components}
         \acro{EE}{energy efficiency}
\end{acronym}

%% file: sections/introduction.tex
\IEEEPARstart{B}{\lowercase{y}} $2034$, global mobile data traffic is expected to grow by \textcolor{black}{five- to nine-fold}, with \ac{AI} accounting for one-third of the traffic. 
There is an industry-wide consensus that $6$G is expected to launch around $2030$ \textcolor{black}{with new spectrum in} \ac{FR3} \cite{10605910}, which spans $7.125$-$24.25$ $\GHz$ \cite{3gpp2021}. To meet \textcolor{black}{the $6$G} timeline, it is crucial to allocate the necessary spectrum a few years in advance, \textcolor{black}{in order to ensure incumbents are accommodated and that the technology is ready for mass adoption at the global launch of $6$G.} Companies are actively conducting proof-of-concept \textcolor{black}{product trials} to address bottlenecks and tackle \textcolor{black}{myriad} implementation challenges. At the same time, regulators must determine how the spectrum will be allocated, and operators need to develop viable business cases \textcolor{black}{and deployment/integration strategies for nationwide $6$G networks.}
\subsection{The FR3 Spectrum}
\begin{figure}[!t]
\centering
\includegraphics[width=3.5in]{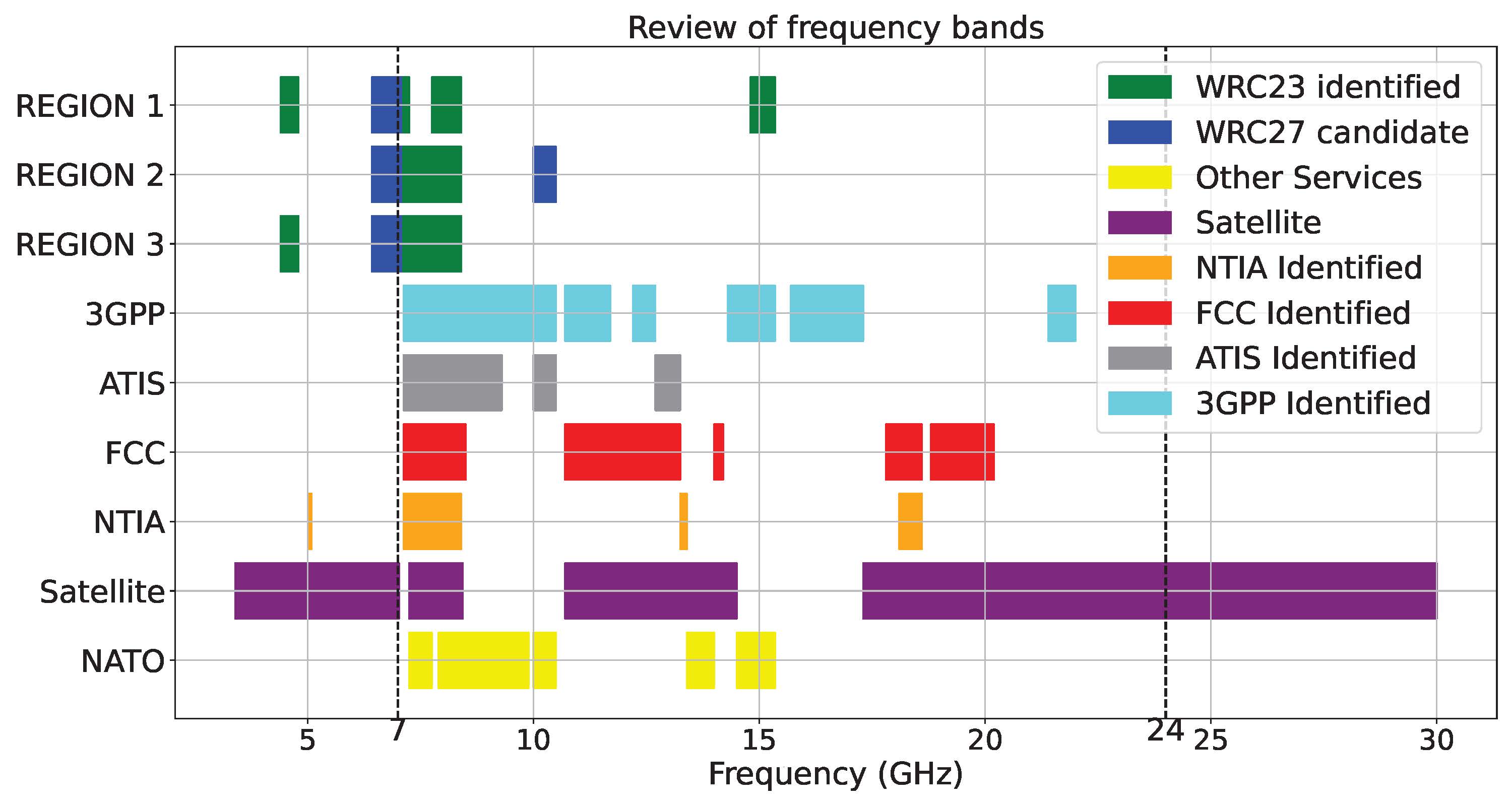}
\caption{The spectrum showing the placement of FR3 \textit{Golden band}, relative to its other counterparts according to different bodies and in various regions \cite{10605910}.}
\label{fig:fig_1}
\end{figure}

 As the global wireless industry advances towards $6$G, the upper mid-band frequencies, being a part of the high \ac{SHF} bands, are critical resources due to their balance between coverage and bandwidth availability, compared to lower \ac{FR1} (up to $7.125 \GHz$) and higher \ac{FR2} ranges, which includes $24.25 \GHz$ to $71 \GHz$ \cite{3gpp2024}. Often termed the \textit{“Golden Band”}, or \textit{"Goldilocks Spectrum"}, for $6$G, FR3 frequencies ($7.125$-$24.25$ $\GHz$ \cite{3gpp2021} as depicted in Fig. \ref{fig:fig_1} \cite{10605910}) are particularly suited for enhancing network capacity while maintaining reasonable propagation characteristics, offering moderate propagation losses, enabling extensive urban and suburban reach using existing towers.
 \textcolor{black}{\input{actions/spectrum-allocated.tex}}
 
The FR3 spectrum holds growing importance and interest for industry and regulatory bodies like the U.S. \ac{NTIA}, the \ac{3GPP} and the \ac{FCC} 
, who are currently evaluating its potential alongside \textcolor{black}{existing mobile radio} bands in order to expand cellular services. Besides mobile operator use, the upper mid-band is being considered to coexist with incumbent services such as satellite communications, radio astronomy, and earth exploration, in addition to warfare activities, such as the next generation jammer mid-band operating within $509 \MHz$ to $18 \GHz$ band. 

Notably, the \ac{NTIA} highlighted specific FR3 bands, including $7.125 \GHz$-$8.4 \GHz$, 
and $14.8$-$15.35 \GHz$, as part of the \ac{ITU} World Radio Conference $2023$ (WRC-$23$) agenda, emphasizing \textcolor{black}{the pending authorization in the USA and hence the} need for precise channel modeling to ensure effective spectrum utilization and sharing.

\subsection{\textcolor{black}{Improvement items by \ac{ITU}, 3GPP \& Organization}}
\label{sec:improvement-items}
\textcolor{black}{\input{actions/improve-itu}}

%% file: actions/spectrum-allocated.tex
U.S. regulators have signalled that $7.25$-$7.4$ GHz will serve as an initial slice for FR3, which is favorable as it avoids the $22$ GHz water-vapour and $60$ GHz oxygen absorption peaks, keeping rain attenuation modest enough for macro-cell coverage. Further FR3 spectrum is expected to be layered outward from this starter band.

%% file: actions/improve-itu.tex
\ac{3GPP} work started in 2024 during Release 19, including \textbf{channel models} for the FR3. The recent FR3 measurements by NYU WIRELESS are unified in Section \ref{sec:channel-char}.
Meanwhile, the \ac{ITU} has identified a list of capabilities  in \ac{IMT}-$2030$ in the form of improvements and additional capabilities compared to those in \ac{IMT}-$2020$. The improvements include: 
a \textbf{peak data rate} of up to $200 \Gbps$, which is $10 \times$ that of \ac{IMT}-$2020$, in addition to \textbf{user experienced data rates}, which are $300 \Mbps$ and $500 \Mbps$, i.e a factor of $3$ to $5$ improvement. The \textbf{area traffic capacity} are expected to be $30$ and $50 \Mbps$ per square meter and the \textbf{spectrum efficiency} was set to an improvement of $1.5$ and $3$ higher than that of \ac{IMT}-$2020$, while \textbf{connection density} is set to a target of $10^{6}$-$10^{8}$ devices per km$^2$.
In Section \ref{sec:itu-enhancements}, we discuss different methods to achieve the aforementioned factors. Moreover, we highlight related challenges for commercial deployment in Section \ref{sec:challenges-commercial}.
Moreover, IMT-2030 also calls for enhancing \ac{RAN}, including flexibility, intelligence, and resiliency, so that a single \ac{RAN} can be sliced, \ac{AI}-optimized, and seamlessly span terrestrial and non-terrestrial links while guaranteeing specific QoS. Besides \ac{RAN}, we mention several open questions and provide potential solutions in Section \ref{sec:open-questions}.

%% file: sections/channel-characteristics.tex
\begin{table*}[htbp]
  \centering
  \caption{Omni PLEs from the CI PL model with 1 m reference distance, Omni DS, and Omni ASA at RX \cite{10605910,Shakya2025,Ying2025UpperMidBand}.}
   \label{table:omnistats}
    \includegraphics[width=\textwidth]{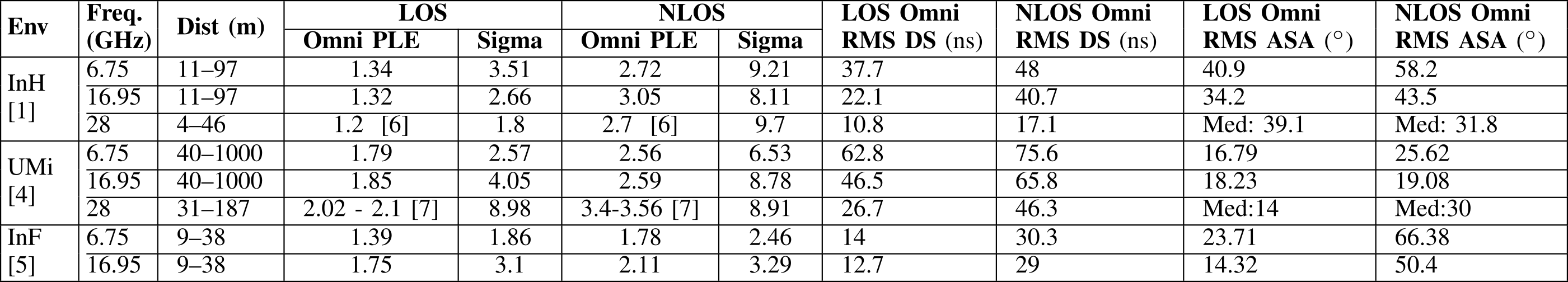}
  \label{table:omnistats}
\end{table*}

\subsection{FR3 Channel Modeling}
\label{sec:FR3-channel-modeling}
The channel model introduced in TR $38.901$ \textcolor{black}{as part of the \ac{3GPP} global standard body} designed to encompass the entire frequency range from $0.5$ to $100$ $\GHz$ \cite{etsi2017138}.
Integrated channel models are indispensable for addressing diverse propagation scenarios, ranging from urban environments to free space, yet \textcolor{black}{\ac{3GPP} developed the TR 38.901 channel models} without many field measurements across \textcolor{black}{much of the} $100 \GHz$ wide swath of the spectrum, \textcolor{black}{and without \textit{any} measurements from the FR3 bands}. \textcolor{black}{Consequently}, the model is an estimate, formulated from sparse measurements across particular bands within the $0.5$ to $100 \GHz$ range.

Channel modeling research for spectra above $6 \GHz$ was completed in Release 17 of TR $38.901$ in April $2022$, offering a comprehensive set of models for evaluating various physical layer technologies. However, the primary focus of $5$G channel modeling has been on frequencies below $6 \GHz$ (FR1) and above $24 \GHz$ \textcolor{black}{(FR3)}. To address this gap, frequency interpolation techniques were used \textcolor{black}{by 3GPP to} estimate channel parameters in the $7$-$24 \GHz$ band. \textit{To establish a more accurate channel model for the FR3 band, it is crucial to validate the TR 38.901 model and carefully examine the details of FR3 channel parameterization.} For this, true empirical characterization of the many channel parameters is required. 


\subsection{\textcolor{black}{FR3 \acp{PLE} over different environments }}
\label{sec:FR3-PLS}
The findings of FR3 channel measurements conducted by NYU WIRELESS in \cite{10605910,Shakya2025, Ying2025UpperMidBand} reveal that for wideband channels at $16.95 \GHz$ in the FR3 band, the omnidirectional \ac{PLE} values, synthesized from directional channel measurements, in \ac{LoS} scenarios were slightly lower than those in \ac{mmWave}, indicating less signal attenuation \textcolor{black}{and more of a waveguide effect} over distance, e.g. the \textit{omnidirectional} \ac{LoS} \acp{PLE} in \ac{InH}, \ac{UMi} and \ac{InF} for $6.75 \GHz$, \textcolor{black}{$16.95 \GHz$}, and $28 \GHz$ are given in Table \ref{table:omnistats}, e.g. 
$1.34$ at $6.75 \GHz$,
$1.32$ at \textcolor{black}{$16.95 \GHz$}, and
$1.2$ at $28 \GHz$ \cite{10605910}. 
In \ac{UMi}, the omnidirectional \ac{LoS} \acp{PLE} are 
$1.79$ at \textcolor{black}{$6.75 \GHz$} \cite{10605910},
$1.85$ at \textcolor{black}{$16.95 \GHz$} \cite{10605910}, and
$2.02$ at $28 \GHz$\textcolor{black}{\cite{Shakya2024TAP}}, 
which \textcolor{black}{indicates slightly less loss over distance than free space channel (e.g., when PLE is $2$) and with less loss at lower frequencies.}
Although antenna patterns and gains vary widely, and higher frequencies enable \textcolor{black}{usage} of directional antennas that can offset channel loss in mmWave bands, 3GPP and researchers often use an omnidirectional channel model to standardize link analysis and ensure consistency in evaluations 
\textcolor{black}{while enabling the use of any type of directional pattern to be applied to the models}
\cite{7417335}, \cite{6175397}.
\textcolor{black}{Omnidirectional} \ac{NLoS} \acp{PLE} were found to be higher than \ac{LoS} scenario as is found in all 3GPP bands as per Table \ref{table:omnistats}, e.g. 
\ac{NLoS} \ac{PLE} for \ac{UMi} of 
$2.56$ at $6.75 \GHz$ \cite{10605910},
$2.59$ at \textcolor{black}{$16.95 \GHz$} \cite{10605910}, and for \ac{InF} the \ac{NLoS} \ac{PLE} is
$1.78$ at $6.75 \GHz$ \cite{Ying2025UpperMidBand}, and
$2.11$ at \textcolor{black}{$16.95 \GHz$} \cite{Ying2025UpperMidBand},
\textcolor{black}{the loss over distance in \ac{NLoS} locations is lower} (\textcolor{black}{e.g., $12$ to $14.5 \dB$ per decade less)} than those observed at \ac{mmWave} frequencies, e.g. \ac{NLoS} \ac{PLE} of \textcolor{black}{$3.4$ to $3.56$} at $28 \GHz$ \textcolor{black}{\ac{UMi} 
 \cite{Shakya2024TAP}}, highlighting how FR3 offers improved coverage in NLOS channels relative to higher-frequency mmWave bands. It is important to note, however, that practical systems use higher antenna gains at higher frequencies, and a common error is to attempt to use \ac{EIRP} to compare coverage at different frequencies, since the transmit and
receive antenna physical apertures are proportionally reduced
with shrinking wavelength if antenna gains are left constant over frequency, such that equal power transmitters with \ac{EIRP} over different frequency will give the false impression that higher bands have far less coverage. In practice, the proper comparison is to consider equal antenna aperture areas at different frequencies, where in practice the extra array gain at mmWave can overcome the additional channel loss by many dB \cite{8269045}.
\textcolor{black}{In this work, Fig. \ref{fig:multi_band_sensing} and Fig. \ref{fig:rate_coverage} do not consider the reality of greater channel gain at higher frequencies, and hence overestimate the achieavable capacity of lower frequencies and  underpredict higher frequency rates.}

Referring to Table \ref{table:omnistats}, 
the \ac{PLE} for \ac{LoS} in \ac{UMi} is $1.79$ for \textcolor{black}{$6.75 \GHz$} and \textcolor{black}{$2.02$ to $2.1$ \cite{Shakya2024TAP}} for $28 \GHz$, implying 
\textcolor{black}{greater coverage with a $2.2 \dB$ stronger signal per decade of distance, resulting in}, \textcolor{black}{$4.4\dB$} over $100\meters$ and \textcolor{black}{$6.6\dB$ stronger signal} over $1\kilometers$, which clearly demonstrates how a lower \ac{PLE} translates into significant coverage improvement, which may be traded for higher capacity for users within a cell, a factor that will be of paramount importance to carriers when deploying $6$G technologies.
\subsection{\textcolor{black}{FR3 \acp{DS}, \acp{AS} \& penetration losses}}
\label{sec:FR3-spread}
The \ac{DS} of a propagation channel has impact on signaling formats, in addition to accuracy for \ac{ISAC} applications such as position location and multi-user synchronization methods.
Measurements in \cite{10605910}, \cite{Shakya2025}, \cite{Ying2025UpperMidBand} show that \ac{RMS} \ac{DS} decreases with carrier frequency, with FR3 exhibiting smaller \ac{DS} values than sub-$6$ $\GHz$, \textcolor{black}{but larger than at \ac{FR2} mmWave frequencies}, e.g. $14 \nanoseconds$ and $12.7 \nanoseconds$ \ac{LoS} omnidirectional \ac{RMS} \ac{DS} at $6.75\GHz$ and $16.95 \GHz$, respectively for \ac{InF}. The range of \ac{RMS} \ac{DS} was found to be $12.7$-$	37.7 \nanoseconds$ across \ac{InF}, \ac{UMi} and \ac{InH}. 
\textcolor{black}{\ac{DS} values} observed in the field indicates that multipath components are closer in time at \ac{FR3}, suggesting limited temporal dispersion which could favor high-speed, low-latency communications and more accurate location and timing methods in dense environments.

Furthermore, the measurements in \cite{10605910} found that higher FR3 frequencies, i.e. \textcolor{black}{$16.95 \GHz$}, exhibited a narrower \ac{RMS} \ac{AS} compared to the higher \ac{FR1} frequency at $6.75 \GHz$, hence indicating fewer, more focused multipath components, which is beneficial for spatial multiplexing as it reduces interference between signal paths and allows more precise beamforming.

Regarding material penetration, losses were consistently higher at $16.95 \GHz$ in FR3 than at lower frequencies, with losses dependent on material type and polarization configuration, confirming the inherent limitations of FR3 in penetrating certain materials, such as low-emissivity (IRR) glass and concrete \cite{10605910}, yet able to allow more penetration that at FR3 mmWave \cite{10605910}.


\subsection{FR3 frequency-dependent features and losses} 
Frequency-dependent propagation features were measured and revealed in \cite{7481506} for both outdoors and indoors empirically discovered in what eventually became the FR3 bands of $28$ and $73 \GHz$.
\textcolor{black}{For modeling the channel path loss over distance, extensive empirical measurements show the} \textcolor{black}{\ac{CIF}} is more suited for indoors \cite{7481506}, \textcolor{black}{and extends} the CI model
\textcolor{black}{which has a physics-based close-in free space reference distance as a leverage point for the slope of the exponential path loss}, \textcolor{black}{while incorporating} the frequency dependence feature of path loss observed indoors.
Perhaps most importantly, rain attenuation and foliage losses vary significantly across the \ac{FR3}.
Following \ac{ITU}-R recommendation (P.$838$-$3$), the specific attenuation at $7 \GHz$ for heavy rain (corresponds to rain rate of $8 \operatorname{mm/hr}$) is \textcolor{black}{$0.04\operatorname{dB/km}$}, whereas at $24\GHz$, the specific attenuation is \textcolor{black}{$ 1.16\operatorname{dB/km}$}, i.e. a gap of \textcolor{black}{$1.16 \dB$} difference over $1 \kilometers$.    
Following Weissberger's model, the foliage loss difference at $24 \GHz$ relative to that at $7 \GHz$ is \textcolor{black}{$14.52 \dB$ at $100 \meters$}. In particular, the foliage loss is \textcolor{black}{$34.66 \dB$} at $7 \GHz$ over $100\meters$, whereas it is \textcolor{black}{$49.18\dB$} at $24 \GHz$ over $100\meters$.   
However, \textcolor{black}{the link} can be improved to extend the transmitter's coverage area via coherent multi-beam combining \cite{6884191}.

\textcolor{black}{It is critical to realize that the 3GPP channel modeling efforts in TR 38.901 \cite{etsi2017138} have up until now ignored the frequency-dependent channel characteristics for most temporal/spatial statistics between $6$-$100$ GHz, hence these real-world data are vital for informing a new version of the 3GPP channel model in coming years \cite{Shakya2025,Ying2025UpperMidBand}.}


%% file: actions/consVspros.tex
\begin{table*}[t]
\centering
\caption{\textcolor{black}{Pros and cons of exploiting the $7$-$24$ GHz FR3 spectrum for $6$G}}
\label{table:pros-cons}
{\color{black}%
\setlength{\tabcolsep}{4pt}
\renewcommand{\arraystretch}{1.05}
\begin{tabularx}{\textwidth}{@{}>{\raggedright\arraybackslash}p{0.48\textwidth}>{\raggedright\arraybackslash}p{0.48\textwidth}@{}}
\toprule
\textbf{Pros / Opportunities} & \textbf{Cons / Challenges \& Research gaps} \\
\midrule
\begin{minipage}[t]{\linewidth}\footnotesize
\begin{itemize}[leftmargin=*]
  \item Wider bandwidth than FR1 but lower pathloss than FR2.
  \item Balances coverage with capacity: outdoor-to-indoor reach that mmWave lacks, but higher throughput than FR1.
  \item The centimeter wavelengths enables large antenna, unlocking extreme \ac{mMIMO}.
  \item Narrower beams than FR1 and richer multipath than FR2 make FR3 attractive for \ac{ISAC}.
  \item Compact high-gain arrays possible on drones. Compact arrays with mutual coupling for tightly coupled colinear antenna arrays can expand bandwidth \cite{10278592}.
  \item Rain/atmospheric loss lower than mmWave.
  \item "Bridge band" for sensing, which helps in mitigating Doppler mismatch between FR1+FR3 or FR3+FR2.
  \item High-resolution ISAC: $100$–$400$ MHz chunks, good for multi-target tracking and imaging.
  \item Multi-band capabilities yielding a diverse view of the environment.
  \item Coverage–capacity sweet-spot: 10 GHz link budget still supports O2I service without mmWave  densification.
\end{itemize}
\end{minipage}
&
\begin{minipage}[t]{\linewidth}\footnotesize
\begin{itemize}[leftmargin=*]
  \item Mid-band inheritance: more loss than FR1, less available bandwidth than FR2.
  \item \ac{PN} at higher frequencies leads to \ac{SNR} degradations.
  \item Aggregating separated sub-bands increases ADC and algorithmic complexity; frequency-dependent channels need characterization.
  \item RF front-end parts (including \acp{PA}, \acp{ADC}, and mixers) covering $7$–$24$ GHz are immature.  
  \item Antenna design for entire FR3 is a challenge.
  \item Unified channel models for terrestrial, NTN and ISAC links still ongoing.
  \item Spectrum coexistence and aggregation policies demand new multidimensional management schemes.
  \item Crowded spectrum: lower FR3 hosts fixed links \& DoD users, in addition to X-band for \ac{SAR}. Upper FR3 overlaps Ku-band. Sharing needed.
  \item Incumbent protection needs accurate sensing of satellite terminals.
  \item Open research gaps: unified near/far-field models, \ac{ELAA}-aware ISAC theory, learning-based spectrum sharing.
\end{itemize}
\end{minipage}
\\
\bottomrule
\end{tabularx}}
\end{table*}

%% file: actions/itu-enhancements-section.tex
\paragraph{Peak data rates \& bandwidth requirements}
Shannon's formula of capacity tells us that doubling the bandwidth offers higher rates than doubling the power.
To keep up with the momentum of different cellular generations, carrier bandwidth should be extended. 
Indeed, \ac{CDMA} in $3$G used a \ac{SB} of $5 \MHz$, whereas \ac{LTE} uses $20 \MHz$. For $5$G \ac{FR1}, the \ac{SB} is $100 \MHz$.
Moving forward, $6$G must open \acp{SB} that are at least $400 \MHz$ wide to meet the increasing capacity demands of applications like \ac{AR}/\ac{VR}. 
For that, if $3$ carriers in \ac{FR3} are supported, one can go up to $1.2 \GHz$ of possible non-contiguous bandwidth. 
This numerology seems very reasonable to keep up with the \ac{ITU} $6$G vision to achieve peak data rates of $50 \Gbps$ to $200 \Gbps$ depending on the scenario.
\textcolor{black}{\input{actions/mapping1.tex}}

\paragraph{Location accuracy}
\textcolor{black}{\input{actions/mapping2.tex}}

\paragraph{Coverage-rate tradeoffs \& Spectrum agility}
\label{subsec:cov-rate-tradeoff}
\textcolor{black}{\input{actions/rate-cover-analysis}}
\paragraph{\ac{mMIMO} \& \Ac{UL} densification}
Doubling the carrier frequency, e.g. $3.5$ to $7$ GHz, introduces a $6$ dB free space pathloss and shorter coherence distances, requiring $N^2$ more half-wavelength elements to fill the same aperture, but this is counterbalanced by an equal $6$ dB antenna gain, while higher frequency NLoS links still face larger penetration and diffraction losses. \ac{mMIMO} can aid in achieving the spectrum efficiency specified in IMT-2030, given a preliminary study on the number of antennas required. 
For IMT-2030 connection density, besides resorting to lower bands for coverage, \ac{UL} densification, where operators can add more \acp{TRP}, already supported in \ac{3GPP} Rel-16, which are physical \acp{BS} within a wireless  system that act as nodes for transmitting and receiving signals to improve  \ac{UL} coverage and capacity by attaining spatial diversity at the macro level so that if one path is blocked, an alternative can be used instead.


%% file: actions/mapping1.tex
A $1024$-\ac{QAM} constellation offers $10 \bpd$. With $12$ antennas providing $12$ spatial \ac{DoF} and $1.2\ \GHz$ overall bandwidth, a peak data rate of $144\ \Gbps$ is achieved. To meet and exceed the IMT-2030 requirement of $200\ \Gbps$, the system can:
\begin{itemize}
	\item Increase spatial streams (e.g., $16$ antennas to offer $192 \Gbps$)
	\item Aggregate more carrier bandwidth additional $400 \MHz $ to achieve $192 \Gbps$.
	\item Combine the above to reach $256\Gbps$, surpassing the IMT-2030 vision of $\pmb{200\Gbps}$.
\end{itemize}

%% file: actions/mapping2.tex
The multiband results in Fig.~\ref{fig:multi_band_sensing} show that scaling carrier frequency and bandwidth by $4 f_{\mathrm{ref}}$ (with $f_{\mathrm{ref}} = 8\ \GHz$) yields a \ac{CRB} for delay estimation below $10^{-3}$ at $17$ dB \ac{SNR}, corresponding to \textbf{sub $\pmb{10}$ cm} location accuracy, fully meeting the IMT-2030 target. When penetration loss at high bands is an issue, hopping to lower frequencies can attain better \ac{SE}.

%% file: actions/rate-cover-analysis.tex
In Fig. \ref{fig:rate_coverage}, we run a coverage-rate analysis for four distinct frequencies at FR3. We have allocated more bandwidth to higher frequencies, i.e. $7$ GHz, $14$ GHz,$18$ GHz and $24$ GHz are allocated $100$ MHz, $200$ MHz, $300$ MHz, and $400$ MHz, respectively.
\textcolor{black}{\input{actions/mapping3.tex}}

%% file: actions/mapping3.tex
Analysis shows that up to $200 \Mbps$, lower frequencies dominate, with $7 \GHz$ sustaining $200\ \Mbps$ for $30\%$ of the time. Beyond $200\ \Mbps$, the optimal band shifts to $18\ \GHz$ (up to $575 \Mbps$), and again at $850 \Mbps$, demonstrating the need for spectrum agility in \ac{FR3}. Dynamically hopping between FR3 frequencies can maintain optimal rate-coverage tradeoffs and deliver IMT-2030 user experienced data rate requirements for $\mathbf{\geq 1\ Gbps}$ peak and high average rates.

%% file: actions/compare-table.tex
\begin{table*}[t]
\centering
  \caption{\textcolor{black}{Comparison among FR1, FR2, and FR3}}
  \label{table:compare}
{\color{black}%
  \small                    
  \setlength{\tabcolsep}{3pt} 
  \renewcommand{\arraystretch}{1.15} 
  \begin{tabular}{|c|c|c|c|c|c|c|c|}
    \hline
    \textbf{FR}  & 
    \textbf{Deployment feasibility} & \textbf{Coverage} &
    \textbf{Rate} & \textbf{Hardware Implications} &
    \textbf{PLE} & \textbf{DS} & \textbf{AS} \\ 
    \hline
    FR1 &  Outdoors      &
    Wide    & Low      & Low cost   & Low & High & Broad \\ \hline
    FR2 &  Indoors \& backhaul &
    Narrow  & High     & High complexity  & High & Low  & Narrow \\ \hline
    FR3 & Outdoors \& indoors &
    Medium  & Mid & Flexible &  Moderate & Moderate & Moderate \\ 
    \hline
  \end{tabular}}

\end{table*}

%% file: sections/commercial-challenges.tex
In this section, we highlight some challenges for commercial deployment supporting FR3, which include
\paragraph{Non-contiguous bandwidth \& dynamic spectrum environment}
Non-contiguous carrier aggregation requires higher resolution \acp{ADC} and \acp{DAC} and more sophisticated baseband algorithms, thereby increasing processing complexity and power consumption. Enabling dynamic spectrum access and coexistence with satellite, radar, or other wireless services operating in FR3 is crucial.

\paragraph{Phase noise (PN) at higher frequencies}
\input{actions/phase-noise}

\paragraph{Spectrum coexistence}  
Lower FR3 bands overlap fixed-link and defense allocations, while upper FR3 intersects Ku-satellite services, necessitating accurate incumbent sensing and multidimensional spectrum-sharing policies.

\paragraph{RF frontend, impairments \& linearity}  
\acp{PA}, mixers, and tunable filters capable of broadband operation over FR3 are not yet available at commercial scale.
 In addition, RF front ends, have to meet stringent adjacent channel leakage power ratios over multiple sub-bands within size constraints of \acp{UE}.
 Moreover, carrier frequency offset and inphase/quadrature imbalance vary across bands, requiring new  estimation and compensation techniques.
 Furthermore, easily tunable filters to operate smoothly across different FR3 bands is a crucial aspect for multi-band operations in order to leverage the full capability offered by \ac{FR3}.
 Also, operating in FR3 necessitates advanced electronic components, such as \acp{ADC} and \acp{PA}, which must support minimal distortion and maintain linear performance over wider bandwidths.

Additional challenges can be found in Table \ref{table:pros-cons}.

%% file: actions/phase-noise.tex
\ac{PN} arises as noise occurring from random and rapid phase fluctuations of a given signal.
Hence, \ac{PN} in both clocks and oscillators  can create impact the performance, especially at higher frequencies. 
In sampling, for a given jitter error, higher frequency signals will suffer from higher phase errors, therefore degrading the clock \ac{SNR}. 
In analog, a given amount of time jitter also leads to higher \ac{PN} at higher frequencies.
Over FR3, an \ac{SNR} degradation caused by \ac{PN} translates to a loss of $10.7 \dB$.

%% file: actions/open-questions.tex
\textbf{Question 1:} \textit{How do recent channel-measurement findings at new FR3 mid-band frequencies affect strategies in 5G-NR (and beyond), and what protocol or metric modifications are being considered within 3GPP to maintain link quality and \ac{EE}?}\\
The newly reported \acp{AS} in \cite{Shakya2025} implies that \textit{modifications are required for beam management in the new goldilocks FR3 mid-band spectrum.
For sensing applications in \ac{ISAC}, this also means that precise and accurate beam alignment is needed for target detection, localization and tracking.}
In addition to channel modeling, the call for new metrics and figures-of-merit, such as the \textit{Waste Factor}, offer potential for \ac{EE} optimization in $6$G wireless systems.

\textbf{Question 2:} \textit{What is an optimal fully digital design that can balance sensing and communication tasks with reasonable power consumption ?}\\
Hybrid beamforming seems like a good tradeoff to alleviate the power consumption problem, by trading-off some beamforming and beam selection gains, which, in turn has negative reverberation effects on the received \ac{SINR} and \ac{SE} for communications, in addition to detection and localization accuracy for sensing. 

\textbf{Question 3:} \textit{How to deal with the difference in coherence time between the lower and upper ends of FR3 ?}\\
The lower end frequencies of FR3 can be allocated to more highly dynamic environments, whereas the higher end to less mobile environments.

\textbf{Question 4:} \textit{How to promote \ac{SE} and agility?}\\
\textcolor{black}{\input{actions/SEgains.tex}}

\textbf{Question 5:} \textit{Are commercial services and malicious users able to harm satellites on FR3?}\\
Yes. Terrestrial networks can unintentionally disrupt satellites via co-channel interference or even collision risk from dense constellations. Even more, malicious users go further, deliberately jamming, spoofing, hacking or physically attacking spacecraft, potentially denying service leading to  seizing control. For that, mitigation should rely on techniques such as interference nulling and \ac{DSS}, yet none offers total immunity. Consequently, robust coordination rules and continuous security monitoring are essential to ensure satellite safety. Also, harmonious operation with incumbents can be attained via dynamic spectrum access with real-time interference nulling beams and incumbent sensing capabilities.


\input{actions/positioning-accuracy}

\textcolor{black}{\input{actions/distinct}}

\textbf{Question 8:} \textit{What key 3GPP and O-RAN innovations enable spectral agile \& spectral intelligence FR3 operation in 6G networks?}
\begin{itemize}
	\item \textbf{Multi-band aggregation}, which stitches together non-contiguous FR3 blocks so radios can treat scattered spectrum as one wide virtual channel, boosting both sensing and data throughput.
	\item \textbf{Dynamic slot configuration} using \ac{AI}, which can reshape frame structures on the fly across multiple bands to hit the sweet spot between latency, reliability, and capacity.
	\item \textbf{Multi-band pilot design}, which compresses and shares pilots intelligently via sparse allocation, in addition to frequency sharing and adaptive density. Multi-Band pilot design can be realized via AI-driven optimization and compressed sensing techniques to minimize redundancy and manage pilots for better efficiency in multi-band environments.
	\item \textbf{Agile timing advance and scalable numerology}, which keep devices tightly synchronized as they hop among carriers with different sub-carrier spacings, sustaining low-latency links.
\end{itemize}
In addition, an AI-guided RAN intelligent controller (RIC) orchestrates carrier aggregation, beamforming, interference mitigation, and power-adaptive O-RUs so the full FR3 spectrum is exploited efficiently and energy-consciously. Near-RT xApps run inside the RIC’s real-time loop, reacting within milliseconds to spectrum-usage shifts to retune carriers and beams, giving the network reflex-like agility.

\textcolor{black}{\input{actions/agility-legacy-ORAN}}
%


%% file: actions/SEgains.tex
Frequency hopping is interesting in \ac{NLoS} with severe blockages emanating from building materials, where gains depend on the hopping strategy. However, to study hopping gains, practical antenna gains based on aperture size that fits a \ac{BS} and \ac{UE} must be used for assessment, as higher band systems employ higher antenna gains than lower bands. Typical additional link budget gains are about $26$ dB at the \ac{BS} and $13$ dB at the \ac{UE} for $28$ GHz, compared to $10$ dB and $6$ dB respectively at $7$ GHz. Using realistic gains, rather than the omni-directional assumptions in TR 38.901, will yield much more reasonable performance curves, often showing only marginal improvements for mid-band compared to high-band in such scenarios.

%% file: actions/positioning-accuracy.tex
\textbf{Question 6:} \textit{How does FR3 propagation characteristics, effective bandwidth, and antenna array size in the FR3 band bound the positioning error of \ac{RSS}-, angle-, and time-based cellular localization with respect to the IMT-2030 $1$–$10$ cm target?}\\
\ac{RSS}-based systems alone, even with several synchronized \acp{BS} may not be enough to satisfy the stringent $1$–$10$ cm target set by IMT under typical high pathloss and shadowing. 
Angle-based positioning with large uniform arrays may achieve the required accuracy in \ac{LoS} but fails in dense urban multipath, revealing strong sensitivity to \ac{AS} in this spectrum.
Therefore, joint multiband processing for time-based and angle-based positioning algorithms may be needed to attain the $1$–$10$ cm target.
Part of future work should be able to assess the exact localization accuracy over FR3 under its diverse propagation characteristics.

%% file: actions/distinct.tex
\textbf{Question 7:} \textit{What new propagation traits, never captured in FR1/FR2, can FR3 reveal?}\\
One fundamental feature of operating in the FR3, particularly relevant for sensing, is the phenomenon of diffuse scattering \cite{8761205}, leading to \ac{DMC}.
Surfaces and objects are seen differently over FR3 depending on the wavelength of the incident wave \cite{8761205}.
Within FR3, and for \ac{ISAC} applications, a target/object interacting with a channel at the lower end of FR3, will naturally have different diffuse scattering properties than that at the upper end of FR3, even when considering the second-order moments of that channel. In turn, the \ac{DMC} statistics can offer a distinctive feature of identifying and classifying different targets across FR3.
In addition, FR3 will accommodate hybrid near-field and far-field beamforming and algorithms as the Fraunhofer distance varies significantly over the lower/upper FR3 spectrum.

%% file: actions/agility-legacy-ORAN.tex
Regarding legacy RAN, even where incumbent non-O-RAN platforms dominate, $6$G spectrum agility can still be achieved.
Operators can for example: 
$(i)$ retrofit wideband and multiband remote radio units and software-defined basebands already supported by most LTE and $5$G sites; 
$(ii)$ activate standards-based features such as \ac{DSS}, expanded carrier aggregation, in addition to integration with other important technologies such as self-organizing networks and auto refarming.  
$(iii)$ In addition, an \ac{AI}-driven, closed-loop assurance layer that ingests real time RAN data, processes it through a machine-learning pipeline, and then issues automated network actions allows carriers to be split, resized, or retuned on the fly whenever slice traffic or service requirements change. 
Such approaches can deliver O-RAN like features without a disruptive replacement of legacy RAN hardware.

%% file: sections/conclusions.tex
This paper outlined a vision for $6$G in the \ac{FR3} spectrum, exploring its upper mid-band potential, key challenges, and necessary standardization changes. We highlighted \textcolor{black}{\ac{FR3}'s} advantages in capacity, coverage, and \ac{SE}, while emphasizing the impact of its channel characteristics on beam management. 
Based on our \ac{FR3} measurements, we underscored the need for frequency hopping to mitigate blockages, maximize \ac{SE}, and incumbent coexistence, reinforcing the importance of spectrum agility. We examined the potential of \ac{mMIMO} for \ac{FR3} and discussed design challenges, particularly in \textcolor{black}{\ac{UL}}. 
For sensing, we analyzed performance bounds for \ac{FR3} \ac{ISAC} multiband sensing. Additionally, we detailed key $6$G standardization features, including \ac{3GPP} radio frame modifications and O-RAN-based innovations, using RIC-driven xApps and rApps for intelligent spectrum management. However, open questions remain regarding FR3 propagation characteristics, PHY design, and regulatory considerations.